\documentclass[prb,twocolumn,amsmath,amssymb]{revtex4}

\usepackage{bm}
\usepackage{ulem}
\usepackage{dcolumn}
\usepackage[dvipdfmx]{graphicx}
\usepackage{color}

\newcommand{\1}{\mbox{1}\hspace{-0.25em}\mbox{l}}
\newcommand{\ket}[1]{|#1\rangle}
\newcommand{\bra}[1]{\langle#1|}

\newcommand{\tr}{{\rm tr}\,}

\begin{document}

\preprint{preprint}

\title{
Disentangled Topological Numbers by a 
    Purification of Entangled Mixed States 
for Non-Interacting Fermion Systems 
}

\author{Takahiro Fukui$^1$\footnote{fukui@mx.ibaraki.ac.jp}}
\author{Yasuhiro Hatsugai$^2$}
\affiliation{$^1$Department of Physics, Ibaraki University, Mito 310-8512, Japan}
\affiliation{$^2$Institute of Physics, University of Tsukuba, 1-1-1 Tennodai, Tsukuba, Ibaraki 305-8571, Japan}

\date{\today}

\begin{abstract}
We argue that the entanglement Chern number proposed recently 
is invariant under the adiabatic
deformation of a gapped many-body groundstate into a {\it disentangled/purified} one, which
implies a partition of the Chern  number 
into subsystems (disentangled Chern number).
We generalize the idea to another topological number, the Z$_2$
Berry phase for a system with particle-hole symmetry, and apply it to
a groundstate in a weak topological phase 
where the Chern number vanishes but the groundstate nevertheless has edge states.
This entanglement Berry phase is especially useful for characterizing random systems with nontrivial 
edge states.
\end{abstract}


\maketitle

A quantum many-body groundstate can be regarded as a mixed state 
if a system is divided into several pieces and  some of them are traced out.
This enables us to define the entanglement entropy  and spectrum 
(Hamiltonian)
\cite{Holzhey1994443,PhysRevLett.90.227902, Ryu:2006fk,PhysRevLett.101.010504}.
These are widely accepted as new tools to characterize  quantum many-body states.
Recently, the entanglement spectrum has been successfully applied to 
topological insulators through the study of   
edge states along the fictitious boundaries between a partition of a system.
\cite{Turner:2010ys,Prodan:2010fj,Cirac:2011ul,Hughes:2011rm,Alexandradinata:2011gf,Huang:2012db,
Fang:2013dq,Lou:2011jk,Tanaka:2012qv}
This may be reflected by the surprising universality of the bulk-edge correspondence.\cite{Hatsugai:1993fk}

Instead of a partition with definite boundaries,
an extensive partition has been introduced: it has been argued
that the entanglement spectrum of a bulk subsystem can be gapless, at which a topological phase transition
occurs.\cite{Hsieh:2014qy,Hsieh:2014jk}
Not only the spectrum of the entanglement Hamiltonian
but also the corresponding eigenstates are useful to characterize the topological phases of the
original model.
We have shown that, taking an extensive but asymmetric partition,
if the entanglement Hamiltonian is 
still gapped, we can define
the entanglement Chern number.\cite{Fukui:2014qv}
The entanglement spectrum and entropy of the bipartition 
reflect the topological properties of the bulk
through the edge states. 
Instead,  
the entanglement Chern number reflects the bulk property directly.
More recently, it has been reported that 
a random partition for a translationally invariant system describes a disorder-driven 
topological transition.\cite{vijay2014entanglement}
Thus, the various kinds of partitioning for a single pure bulk groundstate reveal
its topological properties 
in different environments. 

In this paper,
we argue that the entanglement Chern number and its generalization to other topological numbers
are invariant under the adiabatic deformation to
make the subsystems disentangled. In other words, the original entangled state is eventually modified to 
the single tensor product of the two subsystems.
In this sense, the {\it entanglement topological numbers} may be called 
{\it topological numbers for a disentangled groundstate} or simply
{\it disentangled topological numbers} attached to subsystems.
This also has the meaning of a {\it partition of the topological numbers}. 
The entanglement entropy and spectrum of the bipartition reflect
how the states in subsystems are
entangled in the groundstate wavefunction. 
On the other hand, 
the entanglement topological numbers 
 clarify the property that 
 remains if the entanglement is eliminated or disentangled.
This process of disentanglement may be 
considered as a {\it purification} of the mixed state.
After describing the general idea
   and the validity of the entanglement Chern number, we discuss
the entanglement Berry phases 
applied
to the weak topological (WT) phase,\cite{Fukui:2013mz,Yoshimura:2014qf}
which is topologically nontrivial   in spite of a  vanishing Chern number.

{\it Schmidt decomposition}:
Let $\ket{G}$ be a many-body groundstate of a fermion system,
and let $A$ and $\bar A$ be
a partition of the total system $A+\bar A$. Then, $\ket{G}$ can be Schmidt-decomposed into 
\begin{alignat}1
\ket{G}=\sum_{s,x}D_{sx}\ket{\Phi_s}\otimes\ket{\bar\Phi_x},
\label{SchDec}
\end{alignat}
where $\ket{\Phi_s}$ and $\ket{\bar\Phi_x}$ are, respectively, orthonormal basis states for $A$ and $\bar A$.
The normalization of $\ket{G}$ requires
$\langle G\ket{G}=\sum_{s,x}D_{sx}D^*_{sx}\equiv\tr DD^\dagger=1$.
The singular value decomposition for $D$, $D=U\Lambda V^\dagger$,
where $U_{s\ell}$ and $V_{x\ell}$ are unitary matrices and 
$\Lambda={\rm diag}(\lambda_1,\lambda_2,\cdots,\lambda_m,0,\cdots,0)$ with $\lambda_\ell>0$,
leads to 
\begin{alignat}1
&\ket{G}=\sum_\ell\lambda_\ell|\Phi_\ell)\otimes|\bar\Phi_\ell),
\label{SinValDec}
\end{alignat}
where $|\Phi_\ell)=\sum_s\ket{\Phi_s}U_{s\ell}$ and
$|\bar\Phi_\ell)=\sum_x\ket{\bar\Phi_x}V^*_{x\ell}$.
The number $m$ of nonzero eigenvalues is called the Schmidt number.
The normalization condition for $\ket{G}$ is
$\sum_\ell\lambda_\ell^2=1$.

{\it Reduced density matrix}:
The density matrix of the pure state $\ket{G}$ is $\rho^{\rm tot}=\ket{G}\bra{G}$.
Tracing out $\bar A$ 
and $A$, we have the reduced density matrix
$\rho\equiv\tr\!_{\bar A}\,\rho^{\rm tot}$ in subsystem $A$ and its complementary
density matrix
$\bar\rho\equiv\tr\!_{A}\,\rho^{\rm tot}$ in subsystem $\bar A$ such that
\begin{alignat}1
\rho&=\sum_{s,t}\ket{\Phi_s}(DD^\dagger)_{st}\bra{\Phi_t}
=\sum_{\ell}|\Phi_\ell)\lambda_\ell^2(\Phi_\ell| ,
\nonumber\\
\bar\rho&
=\sum_{x,y}\ket{\bar\Phi_{x}}(D^\dagger D)^*_{xy}\bra{\bar\Phi_{y}}
=\sum_{\ell}|\bar\Phi_\ell)\lambda_\ell^2(\bar\Phi_\ell|.
\label{RedDenMat}
\end{alignat}
The same $\lambda_\ell^2$ appear in these equations
because $D^\dagger D$ and $DD^\dagger$ have the same eigenvalues except for the zero eigenvalue.

{\it Non-interacting fermion system}:
Let $H$ be a Hamiltonian defined on a lattice by
\begin{alignat}1
H=\sum_{i,j}c_i^\dagger h_{ij}^{\rm tot}c_j,
\label{HamDef}
\end{alignat} 
where $i,j$ denote some internal degrees of freedom as well as the sites.
Let $\Psi_{jn}$ be the $j$th component of the $n$th eigenstate of the Hamiltonian 
$h^{\rm tot}$,
$
\sum_jh_{ij}^{\rm tot}\Psi_{jn}=\sum_m\Psi_{im}E_{mn}
$,
where ${E}$ is the energy eigenvalues 
$E={\rm diag}(e_1,e_2,\cdots)$.
The orthogonality and completeness of the eigenstates are expressed by
$\sum_j\Psi^*_{jm} \Psi_{jn}=\delta_{mn}$, 
$\sum_n\Psi_{in}\Psi^{*}_{jn}=\delta_{ij}$, 
or simply $\Psi^\dagger \Psi=\Psi\Psi^\dagger=\1$, if $\Psi$ is regarded as a matrix.
Let us define the normal mode operator $d_n=\sum_j(\Psi^\dagger)_{nj}c_j=\sum_jc_j\Psi_{jn}^*$.
Then, the Hamiltonian is diagonal, 
$H=\sum_ne_nd_n^\dagger d_n$, and the groundstate is given by
$\ket{G}=\prod_{n\le n_F}d_n^\dagger\ket{0}$, where $n_F$ is a state index below which all the states are
occupied.

Let us discuss $\rho$ and $\bar\rho$ in Eq. (\ref{RedDenMat}) for a non-interacting fermion system.
To this end, assume that all the sites and/or internal degrees of freedom 
are divided into two subsystems $A$ and  $\bar A$, 
which have 
$N_A$ and $N_{\bar A}$ dimensions, respectively.
They are denoted by $a,b\in A$ and $\bar a,\bar b\in \bar A$ with 
$a,b=1,2,\cdots N_A$ and $\bar a,\bar b=1,2,\cdots,N_{\bar A}$.
Define
\begin{alignat}1
&\rho=e^{-{\cal H}}/{\cal Z}, \quad \bar\rho=e^{-\bar{\cal H}}/\bar {\cal Z},
\end{alignat}
where ${\cal Z}=\tr\!_{A}\,e^{-{\cal H}}$ and $\bar {\cal Z}=\tr\!_{\bar A} \,e^{-\bar{\cal H}}$.
For the time being, we restrict our discussion to $\rho$.
Since the entanglement Hamiltonian for a non-interacting fermion system is also
non-interacting,\cite{Peschel:2003uq} we set
${\cal H}=\sum_{ab}c_a^\dagger h_{ab}c_b$, where $a,b\in A$.
Let us diagonalize the Hamiltonian $h$ as
$\sum_bh_{ab}\psi_{bn}=\sum_m \psi_{am}{\cal E}_{mn}$, 
where ${\cal E}={\rm diag}(\varepsilon_1,\varepsilon_2,\cdots,\varepsilon_{N_A})$.
Then, introducing the normal mode operator
$f_n=\sum_a(\psi^\dagger)_{na}c_a=\sum_a c_a\psi_{an}^*$, 
we have ${\cal H}=\sum_{n}\varepsilon_nf_n^\dagger f_n$.
$\rho$ is now written as
\begin{alignat}1
\rho&=\frac{e^{-\sum_n \varepsilon_n f_n^\dagger f_n}}{\prod_n(1+e^{-\varepsilon_n})}
\equiv\prod_{n=1}^{N}
\Bigl[
\ket{1_n}\xi_n\bra{1_n}+\ket{0_n}(1-\xi_n)\bra{0_n}
\Bigr],
\label{RhoAProd}
\end{alignat}
where $\ket{1_n}$ and $\ket{0_n}$ are, respectively,  the occupied and vacant states of the 
$n$th fermion defined by $f_n\ket{0_n}=0$ and $\ket{1_n}=f_n^\dagger\ket{0_n}$, and
$\xi_n$ is the Fermi distribution function
\begin{alignat}1
\xi_n
=\frac{1}{e^{\varepsilon_n}+1}.
\label{XiDef}
\end{alignat}
$\xi$, as well as $\varepsilon$, is often called the entanglement spectrum for convenience.
To rewrite $\rho$ in Eq. (\ref{RhoAProd}) in the form of Eq.  (\ref{RedDenMat}), let us define 
the many-fermion state in the occupation number representation 
$\ket{\ell}=\ket{\ell_1\cdots\ell_n\cdots\ell_{N_A}}$ with the occupation number of the $n$th fermion $l_n=0,1$ 
and 
\begin{alignat}1
\lambda_\ell^2=\prod_{n=1}^{N_A} (1-\xi_n)^{1-\ell_n}\xi_n^{\ell_n}.
\label{LamSqu}
\end{alignat}
Then, $\rho$ in Eq. (\ref{RhoAProd}) is now expressed as
$\rho=\sum_{\ell}\ket{\ell}\lambda_\ell^2\bra{\ell}$.

{\it Correlation matrix}:
The two-point correlation matrix is useful as an alternative to the entanglement Hamiltonian.\cite{Peschel:2003uq}
Noting the relation
$c_i^\dagger c_j=\sum_{n,m}\Psi^*_{in}d_n^\dagger \Psi_{jm}d_m$,
we see that the one-particle correlation function is given by
\begin{alignat}1
C_{ij}&\equiv\bra{G}c_i^\dagger c_j\ket{G}
=P_{ji} ,
\label{CorMatDef}
\end{alignat}
where $P_{ji}
=\sum_{n\le n_F}\Psi_{jn}\Psi_{in}^*
=\sum_{n\le n_F}\Psi_{jn}(\Psi^\dagger)_{ni}$ is 
the projection operator to the groundstate.
We can now define the correlation matrices
in subsystems $A$ and $\bar A$ as follows by simply restricting the sites and/or internal degrees of freedom 
in $A$ or $\bar A$:
\begin{alignat}1
&{\cal C}_{ab}\equiv C_{ab}=P_{ba},\quad
\bar{\cal C}_{\bar a\bar b}\equiv C_{\bar a\bar b}=P_{\bar b\bar a} .
\label{CorMat}
\end{alignat}
Alternatively, 
noting the relationship 
$c_a^\dagger c_b=\sum_{n,m}\psi^*_{an}\psi_{bm}f_n^\dagger f_m$,
we obtain
\begin{alignat}1
{\cal C}_{ab}&=\bra{G}c_a^\dagger c_b\ket{G}
=\tr \ket{G}\bra{G}c_a^\dagger c_b
\nonumber\\
&=\sum_{n,m}\psi^*_{an}\psi_{bm}
\left(\tr\!_{A} \,\rho
f_n^\dagger f_m\right)
\equiv \left(\psi\Xi \psi^\dagger\right)_{ba},
\end{alignat}
where $\Xi$ is the diagonal matrix $\Xi={\rm diag}(\xi_1,\xi_2,\cdots,\xi_{N_A})$.
Thus, the $\xi$ are the eigenvalues of ${\cal C}$.\cite{Peschel:2003uq}

The complementary
reduced density matrix  in a  fermionic representation
and the correlation matrix are calculated similarly.
Solving the one-particle eigenvalue equation 
$\sum_{\bar b}\bar h_{\bar a\bar b}\bar \psi_{\bar bn}=\sum_m\bar\psi_{\bar am}\bar{\cal E}_{mn}$,
where $\bar{\cal E}={\rm diag}(\bar\xi_1,\cdots,\bar\xi_{N_{\bar A}})$,
the entanglement Hamiltonian ${\cal H}=\sum_{\bar a,\bar b}c^\dagger_{\bar a}\bar h_{\bar a\bar b}c_{\bar b}$
can be expressed in terms of the normal mode operator $\bar f_n$ $(n=1,\cdots,N_{\bar A})$ as 
$\bar{\cal H}=\sum_{n}\bar\varepsilon_n\bar f_n^\dagger \bar f_n$.
Then, $\bar \rho$ is written in terms of 
$\ket{\bar\ell}$ and the corresponding $\lambda_{\bar \ell}^2$ as
$\bar\rho=\sum_{\bar\ell}\ket{\bar\ell}\lambda_{\bar\ell}^2\bra{\bar\ell}$,
where $\ket{\bar\ell}$ is the occupation number representation of $\bar f_n$ fermions
and $\lambda_{\bar \ell}^2$ is similar to Eq. (\ref{LamSqu}) but with the Fermi distribution function
$\bar \xi_n$ for the energy $\bar \varepsilon_n$.
The correlation matrix is also expressed as
$\bar{\cal C}_{\bar a\bar b}=(\bar \psi\bar\Xi\bar \psi^\dagger)_{\bar b\bar a}$,
where $\bar\Xi={\rm diag}(\bar\xi_1,\cdots,\bar\xi_{N_{\bar A}})$.

It should be noted that, as discussed below Eq. (\ref{RedDenMat}),
$\rho$ and $\bar\rho$ have the same eigenvalues $\lambda_\ell^2$, and hence 
the sets of eigenvalues 
$\{\lambda_\ell\}$ and $\{\lambda_{\bar\ell}\}$ are exactly the same, although the dimensions 
$N_A$ and $N_{\bar A}$
are generically different.
Let us suppose $N_A\le N_{\bar A}$ for simplicity.
Then, this is possible only when $\bar\Xi$ has the same eigenvalues as $\Xi$ except for 0 or 1.
Namely, in an appropriate order, we have
\begin{alignat}1
&\Xi=(\xi_1,\xi_2,\cdots,\xi_{N_A}),
\nonumber\\
&\bar\Xi=(\xi_1,\xi_2,\cdots,\xi_{N_A},\bar\xi_{N_A+1},\cdots,\bar\xi_{N_{\bar A}}),
\label{SpeSma}
\end{alignat}
where the extra eigenvalues $\bar\xi_{N_A+1},\cdots,\bar\xi_{N_{\bar A}}$ are restricted to $0$ and $1$.
We conclude that 
$\bar {\cal C}$ for the larger subsystem $\bar A$
has the same eigenvalues as ${\cal C}$ for the smaller subsystem $A$
plus extra trivial eigenvalues of $0$ or $1$.  

\begin{figure}[h]
\begin{center}
\includegraphics[width=1.0\linewidth]{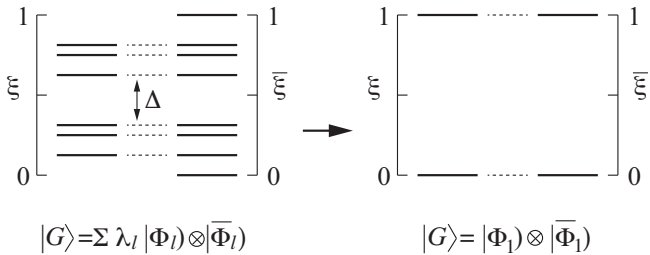}
\caption{
Schematic illustration of a disentanglement deformation.
The spectra of $\xi$ and $\bar\xi$ are identical except for at $0$ and $1$
($\varepsilon=\infty$ and $-\infty$, respectively), and hence
the larger system inevitably has extra eigenvalues of $0$ and $1$. If subsystem $A$ includes
generic eigenvalues $0<\xi<1$, (i) the other one also includes the same generic eigenvalues, (ii)
each reduced density matrix satisfies the grand canonical ensemble with a finite weight 
given by Eq. (\ref{XiDef}),
and therefore, (iii) the groundstate $\ket{G}$ is entangled in the sense that it is composed of multiple
tensor products of the wavefunctions of subsystems $A$ and $\bar A$. If we can deform the spectrum
in the left panel into the spectrum in the right one 
(as if we could take the ``zero temperature limit''),
$\ket{G}$ can be a single tensor product.
This process can be considered as a purification of  the mixed state to the pure state, 
which results in 
a disentanglement of the groundstate wavefunction.  
}
\label{f:fig1}
\end{center}
\end{figure}

Suppose
that the spectrum of the entanglement Hamiltonian  $\xi$ has a gap
at $\xi=1/2$, as illustrated in Fig. \ref{f:fig1},
and that we calculate some topological numbers of the upper bands.
In a generic spectrum, the groundstate is a linear combination of the tensor product in the 
Schmidt decomposition as in Eqs. (\ref{SchDec}) and (\ref{SinValDec}).
Then, suppose that we deform the spectrum adiabatically, making the gap  larger, 
and that we eventually reach an extreme spectrum with $\Delta=1$, i.e., 
all states have $\xi=1$ or 0. 
In this case, the largest eigenvalue, $\lambda_1$, in the singular value 
decomposition $\{\lambda_\ell\}$ becomes $\lambda_1=1$
and the other eigenvalues are $0$, implying that the groundstate is a single tensor product.
Therefore, this adiabatic process can be considered as a
{\it disentanglement} deformation of the 
groundstate wave function between subsystems $A$ and $\bar A$.
From the point of view of  the reduced density matrix, the process is
  considered as a
purification of the mixed state to the pure state.
On the other hand, topological numbers calculated using the eigenstates of ${\cal C}$ and $\bar{\cal C}$
are expected to be invariant in this process since the 
gap between the upper and lower bands never closes.
Therefore, such topological numbers, {\it referred to as entanglement topological numbers},
reveal the topological properties that are  invariant even if the entanglement between 
$A$ and $\bar A$ is eliminated. In this sense, they may  alternatively be called
{\it topological numbers of a disentangled groundstate} or simply
{\it disentangled topological numbers}.
If the groundstate can be represented by a single tensor product such that 
$\ket{G}=|\Phi_1)\otimes|\bar\Phi_1)$, 
a topological number, such as the first Chern number or the Berry phase of $\ket{G}$,
is the sum of the topological numbers of $|\Phi_1)$ and $|\bar\Phi_1)$.
This is indeed possible since $\lambda_1=1$ in Eq. (\ref{SinValDec}). Otherwise,
for generic nonintegral $\lambda_\ell$,
it may be difficult to define integral topological numbers simultaneously 
for $\ket{G}$, $|\Phi_\ell)$, and $|\bar\Phi_\ell)$. 
This also implies that a set of entanglement topological numbers for $A$ and $\bar A$
may be referred to as {\it a partition of a topological number},
provided that the bulk gap of $h^{\rm tot}$ remains open in the disentanglement deformation. 
This can be checked 
by the natural sum rule that the topological number 
of the groundstate is the sum of the two entanglement topological numbers.
Note that  assuming  a finite gap for the
  entanglement Hamiltonian is 
  in contrast to the case with edge states for a bipartition, where
  gapless modes of the entanglement Hamiltonian mainly contribute to the
entanglement entropy.

{\it Translationally invariant system}:
So far, we have used the subscripts $i,j$ for some internal degrees of freedom as well as the sites.
We next consider a system with translational invariance. To this end, we
replace $i,j\rightarrow i\alpha,j\beta$, where $i,j$ and $\alpha,\beta$ denote the sites and species, 
respectively. 
On the $N^d$ lattice in $d$ dimensions  with the periodic boundary condition, 
the fermion operator is now denoted by $c_\alpha(j)$ and its Fourier transformation is 
$c_\alpha(j)=\frac{1}{\sqrt{V}}\sum_k e^{ik\cdot j}c_\alpha(k)$,
where $V=N^d$ and $k_\mu=2\pi/N\times\mbox{integer}$.
For a translationally invariant system,  the Hamiltonian given by Eq. (\ref{HamDef})
becomes $h_{ij}^{\rm tot}\rightarrow h_{i\alpha,j\beta}^{\rm tot}=h_{\alpha\beta}^{\rm tot}(i-j)$
and its Fourier transformation is given by 
$h^{\rm tot}(i-j)=\frac{1}{V}\sum_k e^{i k\cdot (i-j)}h^{\rm tot}(k)$.
Then, the total Hamiltonian is separated into $k$ sectors, 
$H=\sum_k\sum_{\alpha\beta}c_\alpha^\dagger(k)h^{\rm tot}_{\alpha\beta}(k)c_\beta(k)$.
The Schr\"odinger equation for a given $k$ is given by
$\sum_\beta h^{\rm tot}_{\alpha\beta}(k)\Psi_{\beta n}(k)=\sum_m\Psi_{\alpha m}(k)E_{mn}(k)$.
We assume that the groundstate is insulating and that the fermions are occupied up to the $n_F$th band,
$\ket{G}=\prod_{n\le n_F} \prod_{k} d_n^\dagger(k)\ket{0}$,
where the normal mode operators are defined by
$d_n(k)=\sum_\alpha c_\alpha(k) \Psi_{\alpha n}^*(k)$.
The correlation matrix in Eq. (\ref{CorMatDef}) is then
\begin{alignat}1
C_{\alpha\beta}(j,j')
&=\frac{1}{V}\sum_k
e^{ik\cdot(j'-j)}P_{\beta \alpha}(k) ,
\label{CorMatMom}
\end{alignat}
where 
$
P_{\beta \alpha}(k)
=\sum_{n\le n_F}\Psi_{\beta n}(k)\Psi^*_{\alpha n}(k)
$
is the projection operator to the groundstate at a fixed $k$.

{\it Example 1; Entanglement Chern number}:
A typical example of the entanglement Chern number
is the entanglement {\it spin} Chern number \cite{Fukui:2014qv} for the Kane-Mele model. \cite{Kane:2005fk}
The Hamiltonian $h^{\rm tot}(k)$ is given by a $4\times4$ matrix due to 
the spin and the bipartite lattice.
The Rashba term mixes the spins, so that it is basically impossible to define the
spin Chern number simply in the momentum space. \cite{PhysRevLett.97.036808,Fukui:2007sf}
However, projecting
the $4\times4$ $P_{\alpha\beta}(k)$ matrix
in Eq. (\ref{CorMatMom}) into each spin sector $\sigma=\uparrow,\downarrow$ such that 
$P_{\alpha\beta}\rightarrow P_\sigma P_{\alpha\beta}P_\sigma$, where $P_\sigma$ stands for the projection
to spin $\sigma$, 
we have successfully computed 
the set of entanglement Chern numbers $(c_\uparrow,c_\downarrow)$, 
which indeed describes the spin Hall phase when $(c_\uparrow,c_\downarrow)=\pm(1,-1)$.
\cite{Fukui:2014qv}
Although spin is not conserved in a topological insulator
  in general and the time-reversal symmetry guarantees the vanishing of
  the Chern number, disentanglement between the spins
  implies a nontrivial entanglement spin Chern number, which
  justifies the existence of nontrivial spin edge states characterizing the
  phase.
  The advantage of the topological characterization here is that the idea of
  the disentanglement/purification of the mixed state
  to the pure state is simply extended to correlated electrons with an interaction.

\begin{figure}[h]
\begin{center}
\includegraphics[width=0.9\linewidth]{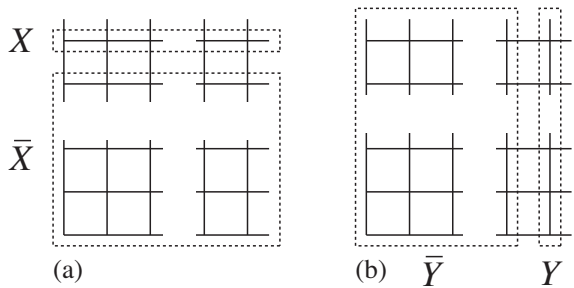}
\caption{
Partition $A$ and $\bar A$ in the case of $n_A=1$, 
(a) for $A=X$ and (b) for $A=Y$.
}
\label{f:Lat}
\end{center}
\end{figure}

{\it Example 2; Entanglement Berry phase}:
The Berry phase here means a winding number for a one-dimensional system. We apply it to a 
two-dimensional system, based on the method in Refs. \cite{Ryu:2002fk,doi:10.1143/JPSJ.75.123601},
to study the nontrivial edge states in a WT phase. 
\cite{Fukui:2013mz,Yoshimura:2014qf}
Consider an $N\times N$ square lattice with the periodic boundary condition.
Let $A$ be a subsystem with
$n_A$ ladders.
The remaining subsystem is denoted as $\bar A$ and is composed of $n_{\bar A}=N-n_A$ ladders. 
For the partitions shown in Figs. \ref{f:Lat}(a) and \ref{f:Lat}(b), 
we set $A=X$ and $A=Y$, respectively.
Let us consider the case $A=X$. 
Since the translational invariance in the $y$ direction 
is broken, 
we regard $j_y$ as the species, and thus the correlation matrix is denoted by
\begin{alignat}1
C_{j_y\alpha,j_y'\beta}&(j_x,j_x')
= \frac{1}{N}\sum_{k_x}e^{ik_x(j_x'-j_x)}
C_{j_y\alpha,j_y'\beta}(k_x),
\end{alignat}
where
\begin{alignat}1
C_{j_y\alpha,j_y'\beta}(k_x)=
\frac{1}{N}\sum_{k_y}e^{ik_y(j_y'-j_y)}P_{\beta\alpha}(k) .
\label{kyCorFun}
\end{alignat}
When $j_y$ and $j_y'$ are restricted within $1\le j_y,j_y'\le n_X$ $(n_{\bar X})$, 
the above correlation matrix is  ${\cal C}$ $(\bar{\cal C})$.
We assume that the eigenvalues $\xi_n(k_x)$ of ${\cal C}_{j_y\alpha,j_y'\beta}(k_x)$
have a spectral gap, as shown in Fig. \ref{f:fig1}. 
This is possible in general for an extremely asymmetric partition with $X\ll\bar X$.
Let $\psi_+(k_x)\equiv(\psi_{j_y\alpha,n_1}(k_x),\psi_{j_y\alpha,n_2}(k_x),\cdots)$ 
be the set of eigenstates with eigenvalues
$\xi_n(k_x)>1/2$.
Then, the entanglement Berry phase for $X$ is calculated by
$\gamma_X={\rm Im}\log\left[\prod_{k_x} {\cal U}_x(k_x)\right]$,
where the U(1) link variable is defined as
${\cal U}_x(k_x)\equiv\det \psi_+^\dagger(k_x)\psi_+(k_x+\delta k_x)$ 
with $\delta k_x=\frac{2\pi}{N}$ a unit of the 
discrete momentum.
Likewise, solving the eigenvalue equation for $\bar{\cal C}$ and/or choosing $R=Y$,
we obtain the other entanglement Berry phases $\gamma_{\bar X}$, $\gamma_Y$, $\gamma_{\bar Y}$.
The sets of $(\gamma_X,\gamma_{\bar X})$ and $(\gamma_Y,\gamma_{\bar Y})$ are considered as
a real-space partition of the conventional Berry phases $\gamma_x(k_y)$ and $\gamma_y(k_x)$,\cite{Ryu:2002fk}
as discussed below.
While the latter are already partitioned
into each $k_y$ and $k_x$ for the pure model, 
the former partition is advantageous when
we study disordered systems.

In what follows,
we consider an anomalous Hall effect model of the Wilson-Dirac type 
\cite{Bernevig:2006fk,Qi:2008fk,Jiang:2014fk,Diez:2014xy,Diez:2014nr}
with an anisotropic Wilson term. The pure model is defined by 
\begin{alignat}1
h^{\rm tot}(k)&=t\sigma_1\sin k_x+t\sigma_2 \sin k_y
\nonumber\\
&+\sigma_3\left[m-b_x(1-\cos k_x)-b_y(1-\cos k_y)\right] .
\label{WilDir}
\end{alignat}
\begin{figure}[h]
\begin{center}
\begin{tabular}{cc}
\includegraphics[width=0.4\linewidth]{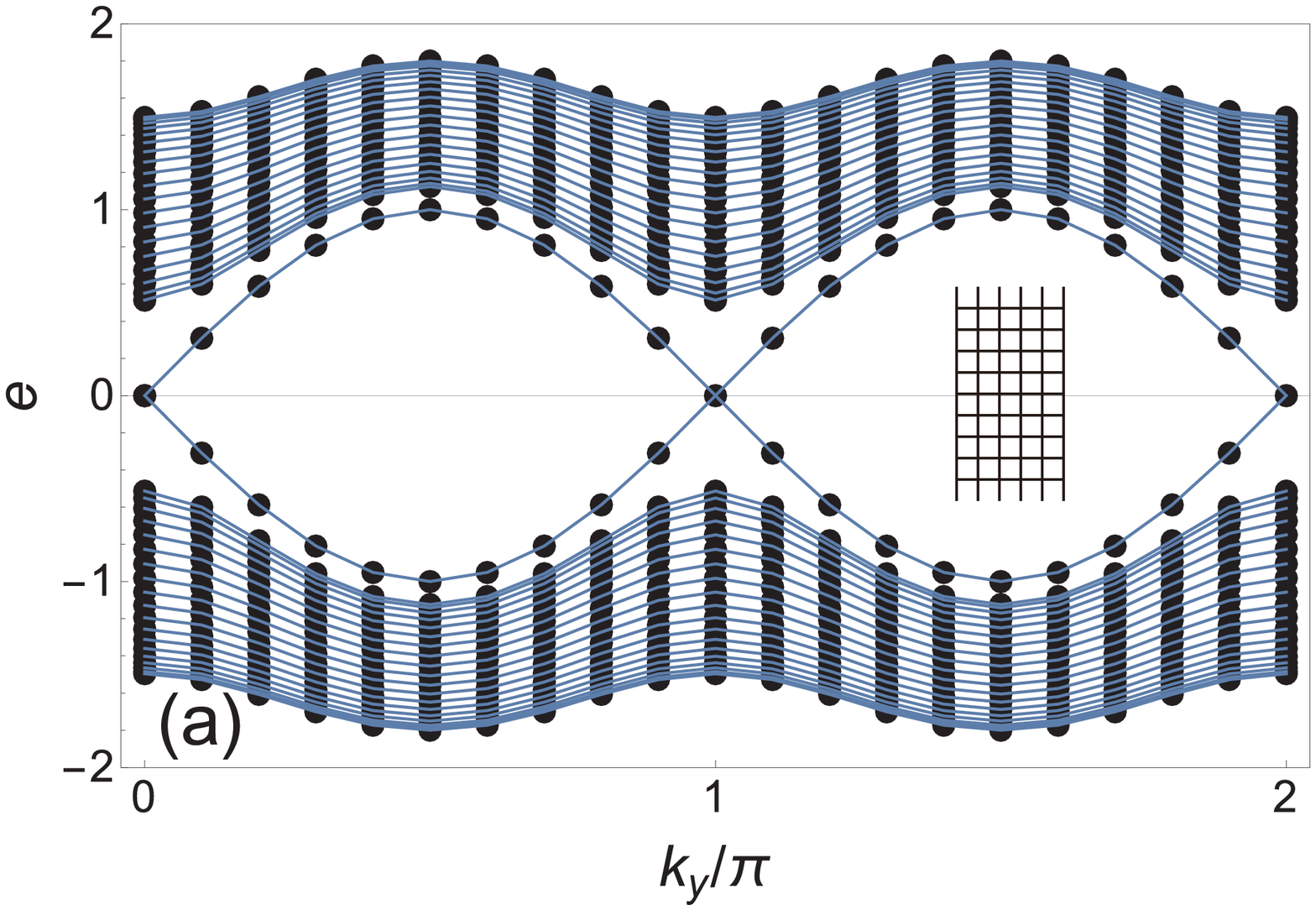}&
\includegraphics[width=0.4\linewidth]{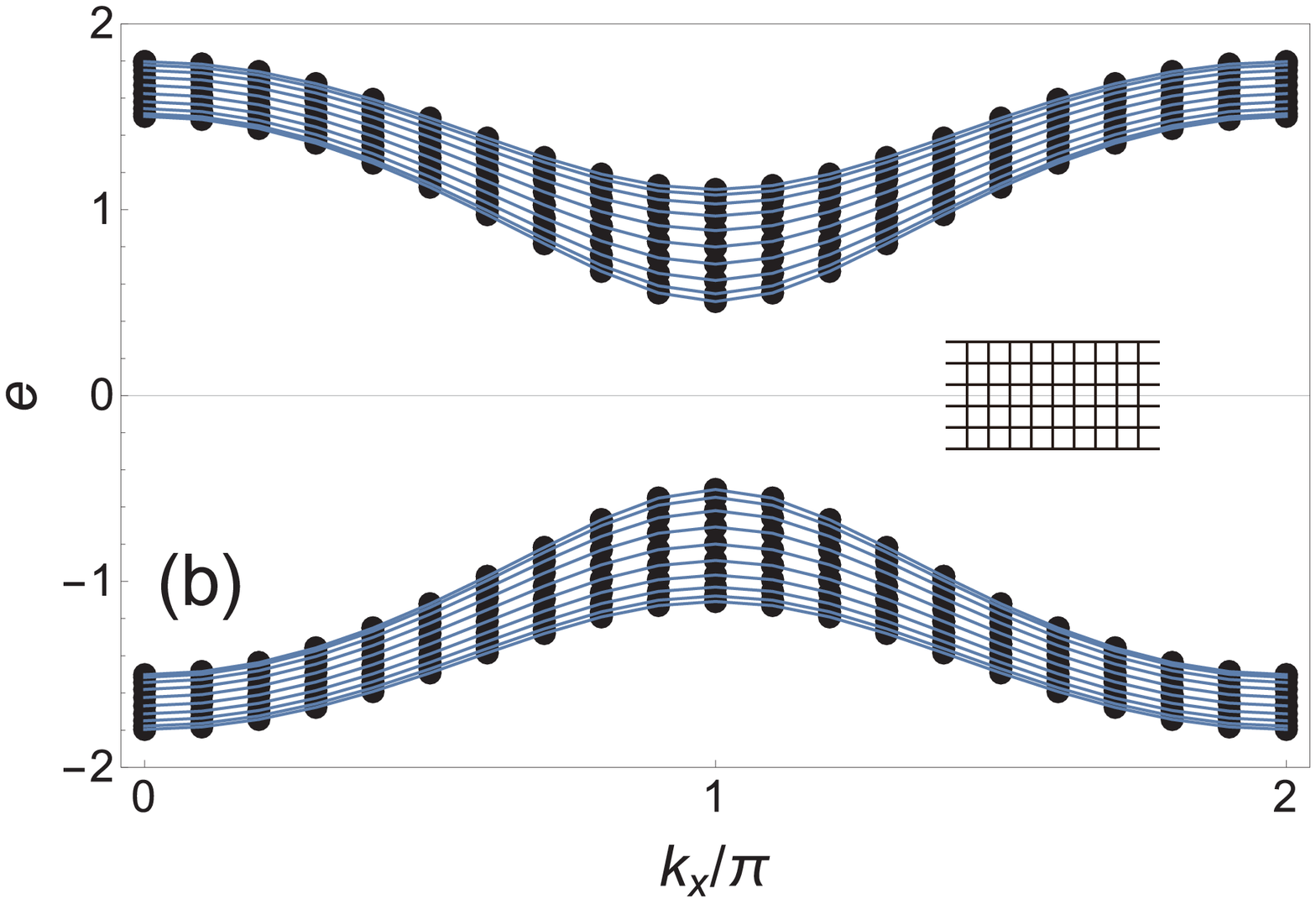}\\
\includegraphics[width=0.4\linewidth]{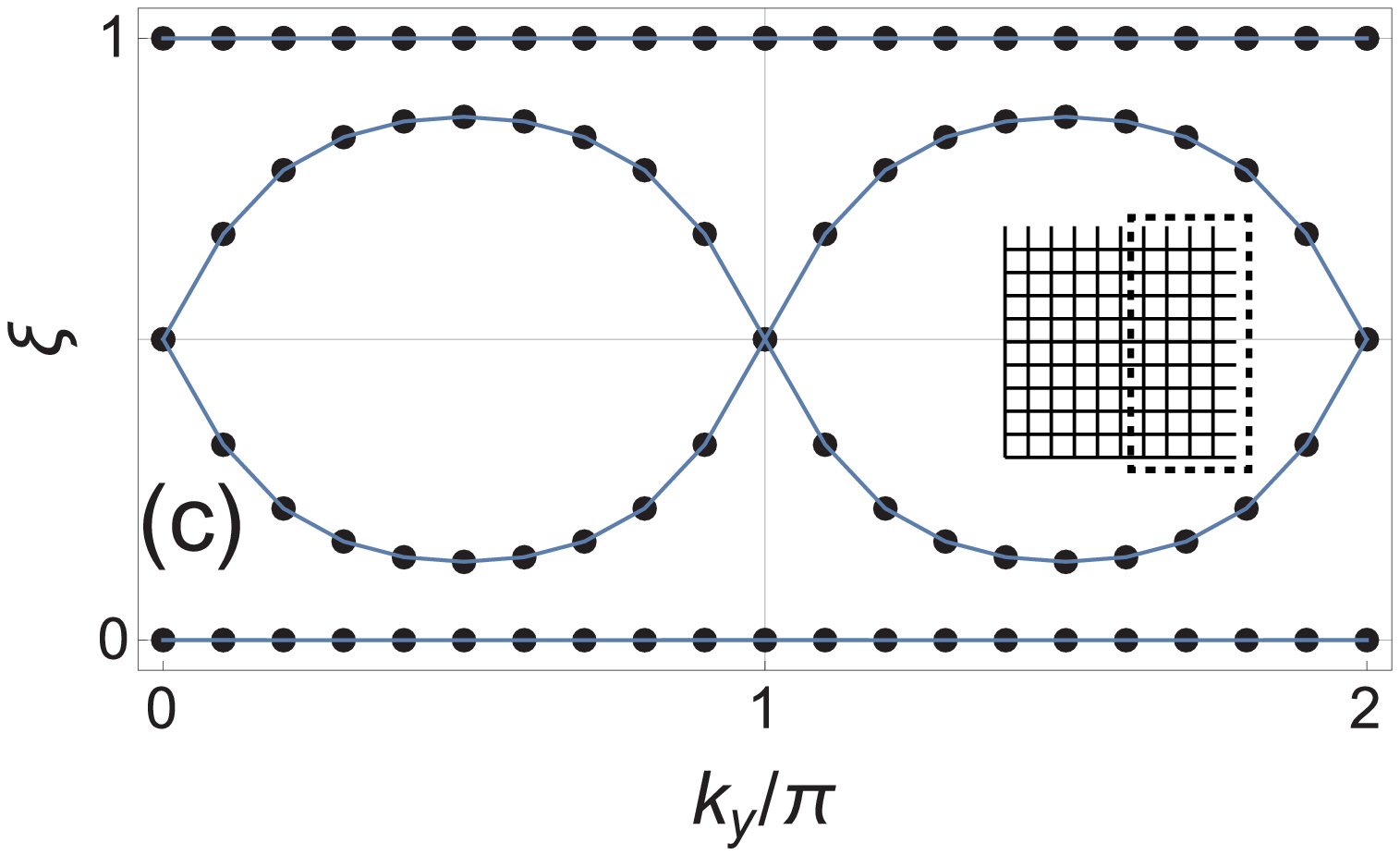}&
\includegraphics[width=0.4\linewidth]{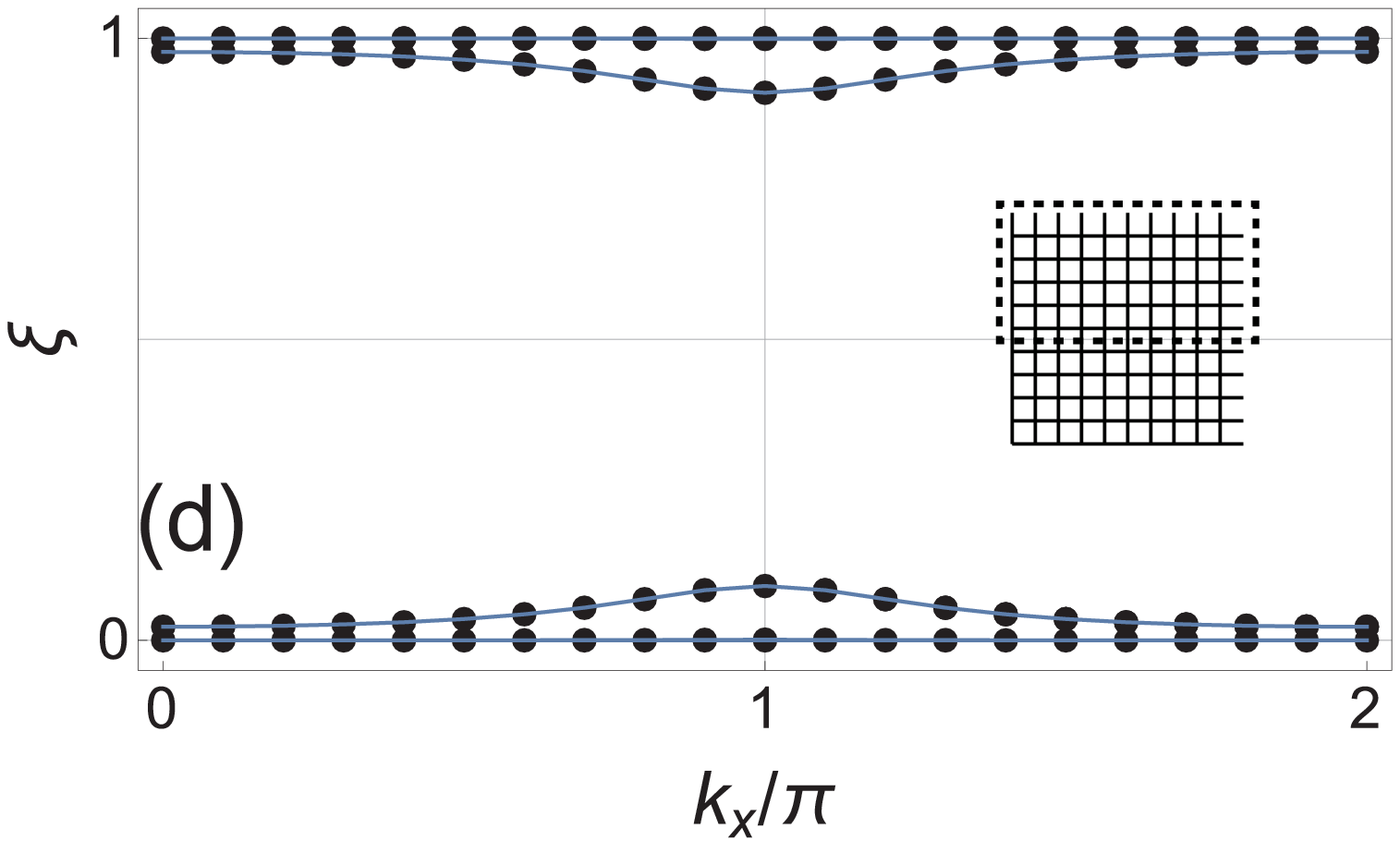}\\
\includegraphics[width=0.4\linewidth]{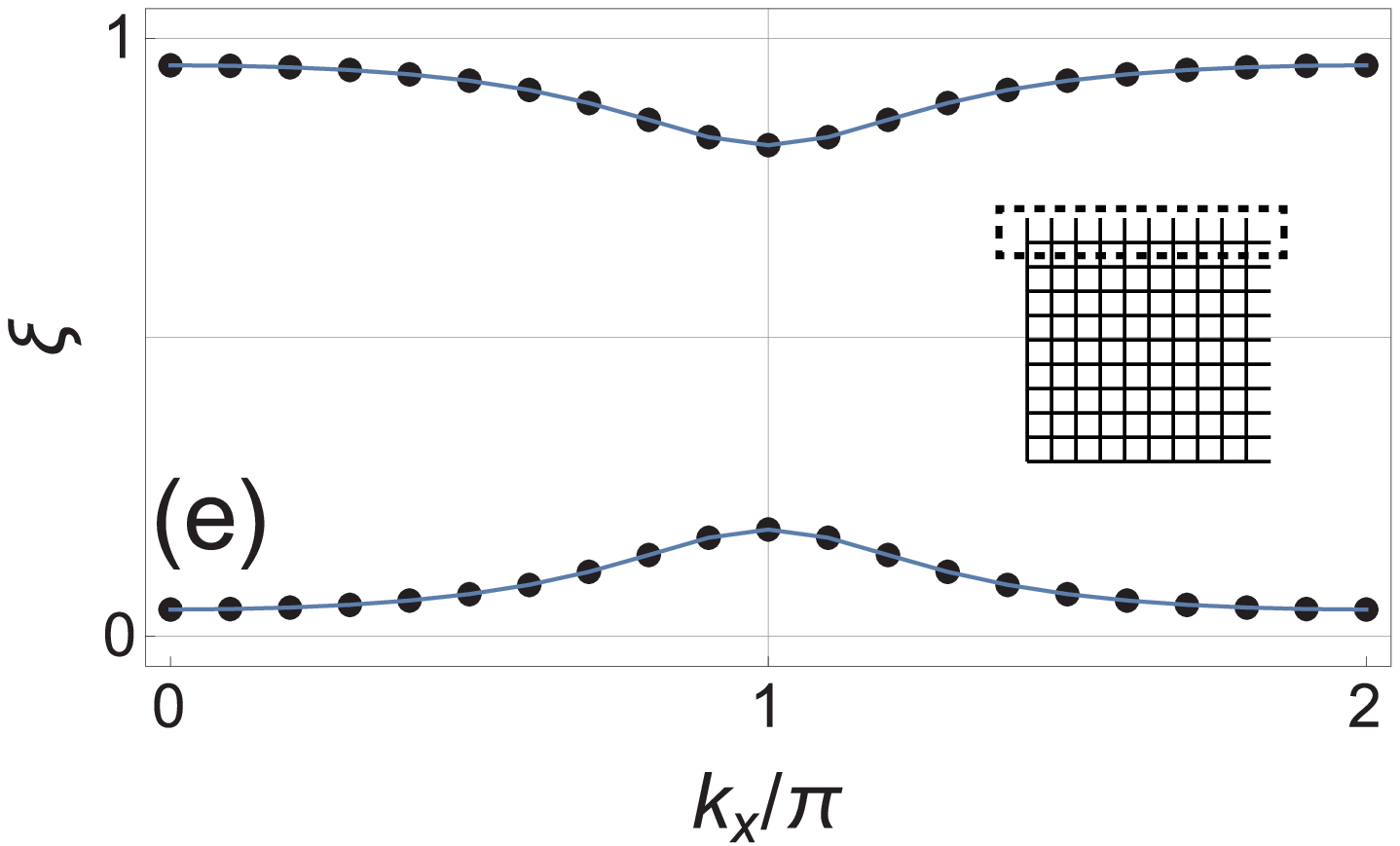}&
\includegraphics[width=0.4\linewidth]{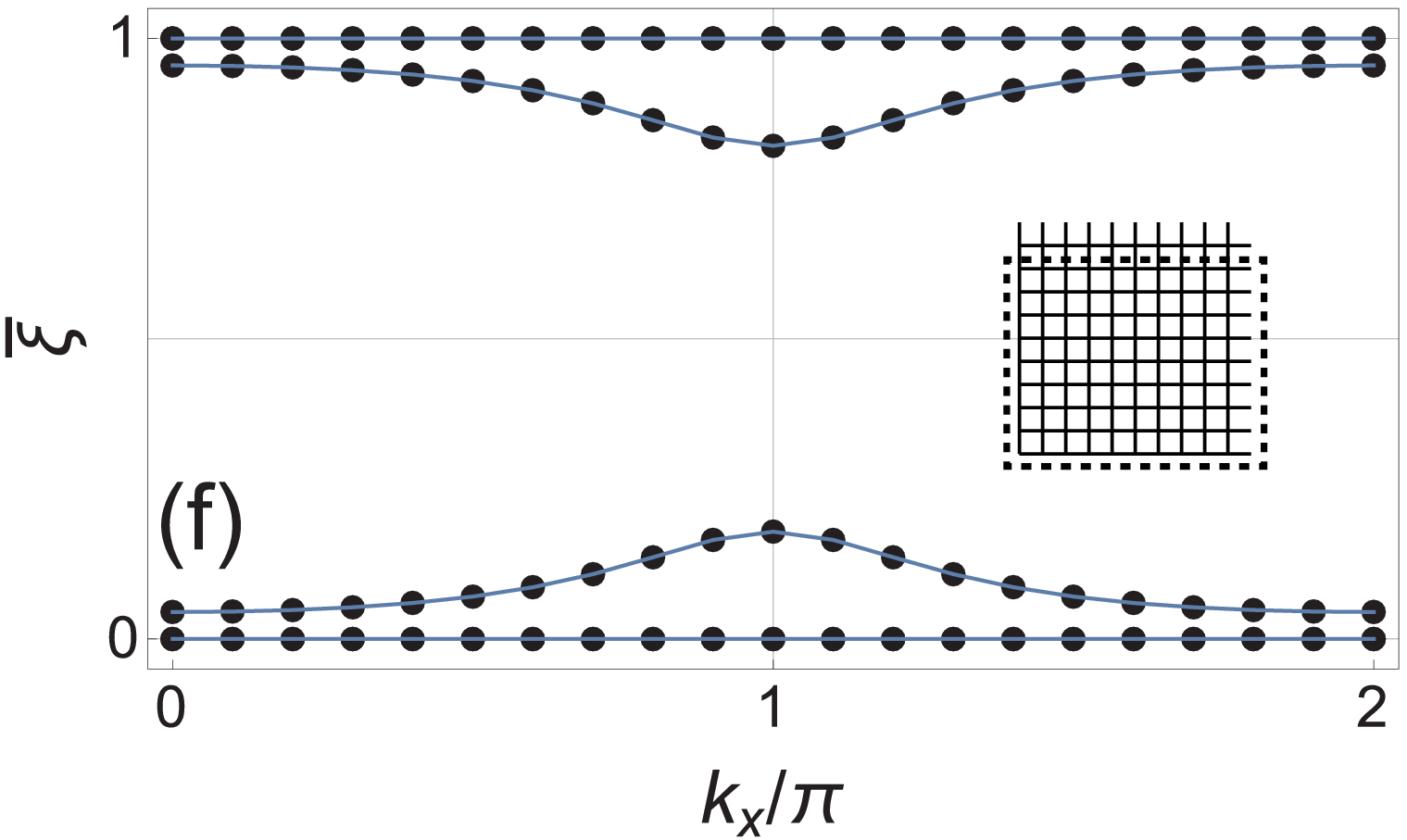}\\
\includegraphics[width=0.4\linewidth]{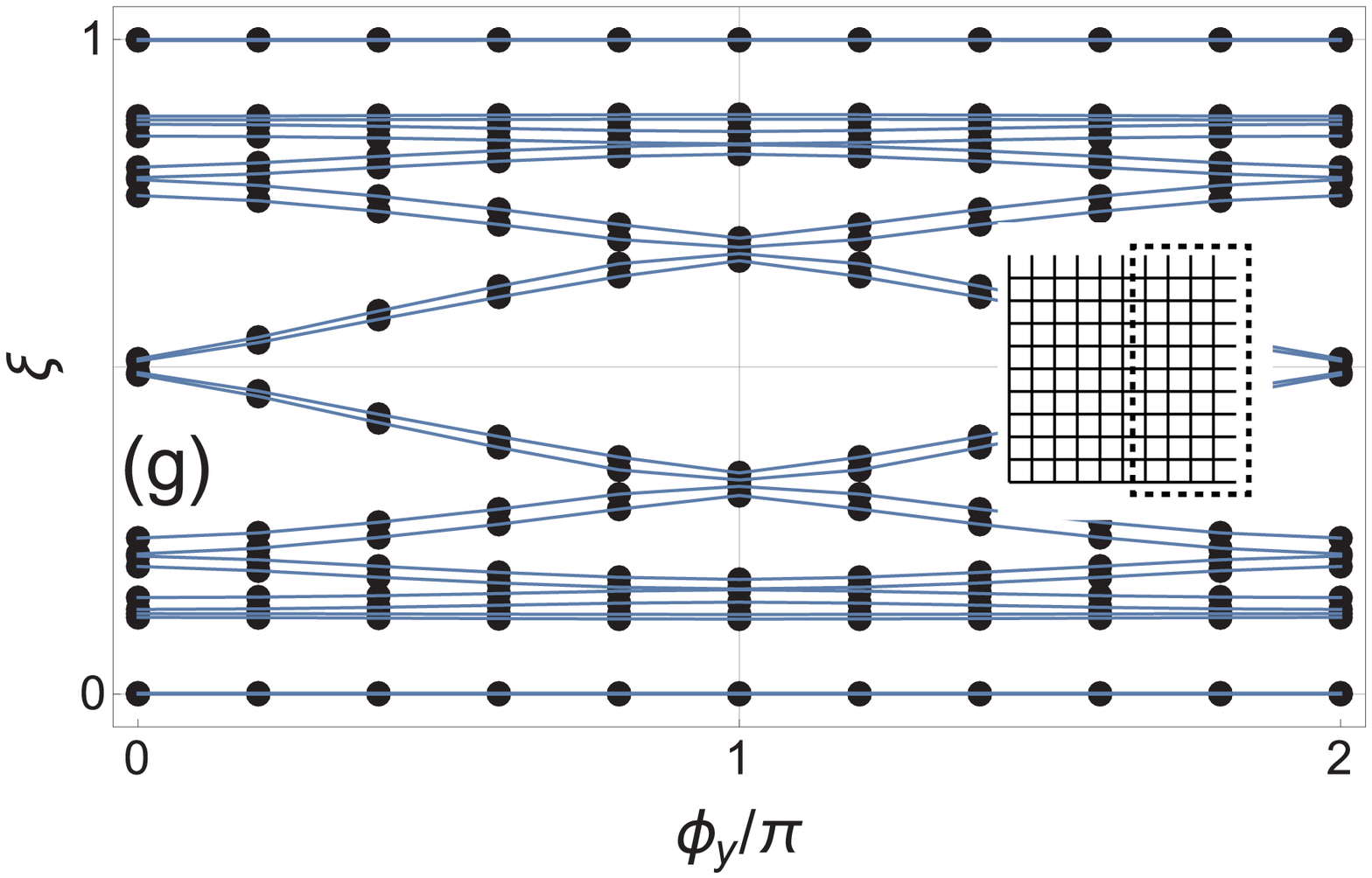}&
\includegraphics[width=0.4\linewidth]{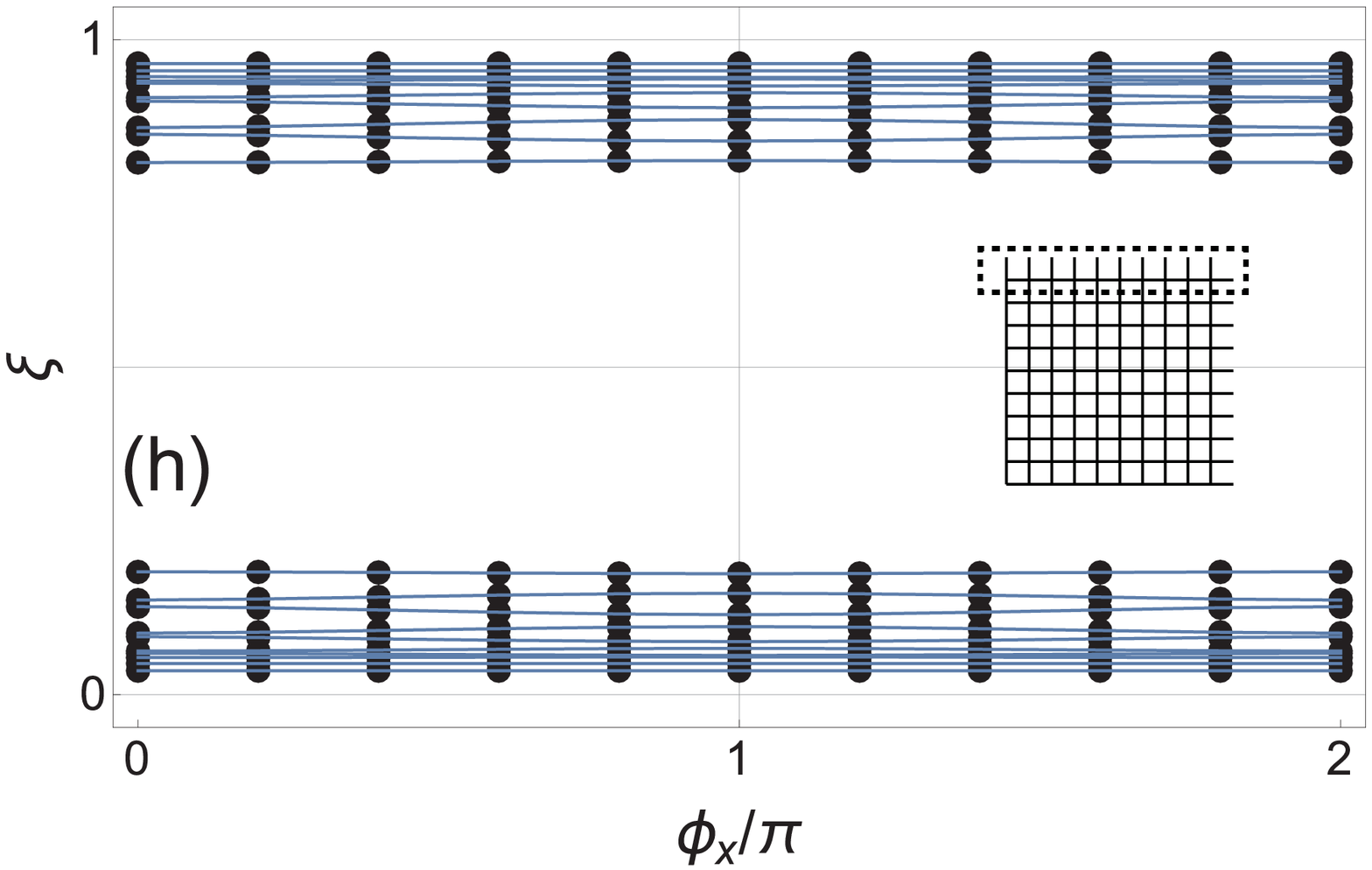}\\
\end{tabular}
\caption{
Various spectra of the model given by Eq. (\ref{WilDir}) 
with $t=1$, $m=1.5$, $b_x=1$, and $b_y=0$ belonging to the $c=0$ WT phase.
Insets show schematic illustrations of the boundary or partition of the system.
The top two figures are spectra of $h^{\rm tot}$ on a cylinder with boundaries (a) parallel to
 the $y$-axis
and (b) parallel to the $x$-axis.
The next two are the spectra of ${\cal C}$ 
(c) for the subsystem $Y$ with $n_Y=N/2$ ($N=20$) and 
(d) for 
$X$ with $n_X=N/2$. The edge states in these figures match the top two figures.
The third two are the same spectra as in (d) but with (e) $n_X=1$ and (f) $n_{\bar X}=19$.
In (f), 18 states are degenerate at $\bar \xi=0$ and $1$. 
The bottom two are those for
the disordered model with the same parameters as above. 
Randomness is included in the mass and hopping such that
$m_j=m+\delta m_j$, $t_{j,\hat \mu}=t+\delta t_{j,\hat \mu}$
with a random distribution $\delta m_j,\delta t_{j,\hat\mu} \in [-0.3,0.3]$. 
(g) Spectrum of ${\cal C}$ for $Y$ with $n_Y=N/2$ ($N=10$) as a function of the twist angle 
$\phi_y$ and 
(h) spectrum of ${\cal C}$ for $X$ with $n_X=1$ as a function of the twist angle
$\phi_x$.
}
\label{f:WTP}
\end{center}
\end{figure}
This model has particle-hole symmetry and its groundstate is characterized by the Chern number. 
In case of an anisotropic Wilson term, 
an interesting $c=0$ phase appears, which 
has edge states. 
This phase has been referred to as the WT phase.  \cite{Fukui:2013mz,Yoshimura:2014qf}
In Figs. \ref{f:WTP}(a) and \ref{f:WTP}(b), we show the spectrum of the system on a cylinder 
belonging to the WT phase, in which edge states can be seen only in (a).  
The question is their stability:
Unless there are specific reasons, 
such states are expected to be unstable against, for example, disorder or interactions.

For a model with particle-hole symmetry, the Berry phase can serve as a Z$_2$ topological invariant.
\cite{Ryu:2002fk}
Furthermore, with translational invariance, 
the conventional Berry phase 
(winding number along $k_x$) $\gamma_x(k_y)$ can be computed.
Then, at a certain $k_y$ where the particle-hole symmetry is enhanced to chiral symmetry,
$\gamma_x$ is quantized as $0$ or $\pi$. 
The Berry phase $\gamma_x=\pi$ for a periodic system is a topological invariant for 
a zero-energy state localized at the end of a finite chain. This state forms, in turn, an edge state
at the boundary parallel to the $y$-axis.
Namely, from the $k_y$-resolved Berry phase $\gamma_x(k_y)$, 
we can predict the edge states 
at the boundary parallel to the $y$-axis.
The Berry phase $\gamma_x(k_y)$ corresponding to Fig. \ref{f:WTP}(a) becomes $\pi$ at $k_y=0$, $\pi$.
However, if the system breaks translational symmetry, $\gamma_x(k_y)$ is no longer defined.
This is a part of our motivation for proposing the entanglement Berry phase.

Let us start with a model with translational symmetry.
In Figs. \ref{f:WTP}(c) and \ref{f:WTP}(d), we show the entanglement spectra for the symmetric partition 
$n_A=n_{\bar A}=N/2$.
It turns out that the edge states in the real space are well simulated by the entanglement spectrum
in a cylindrical partition.
To study the stability of these states in (a) and (c), we calculate 
the entanglement Berry phase
for a minimum subsystem $X$ with $n_X=1$ 
and its complement $\bar X$ with $n_{\bar X}=N-1$, whose spectra are given 
in Figs. \ref{f:WTP}(e) and \ref{f:WTP}(f), respectively.
These are indeed gapped, and the entanglement Berry phase is therefore well defined.
We obtain $(\gamma_X,\gamma_{\bar X})=(\pi,\pi)$ numerically. This is in sharp contrast to the trivial
$c=0$ state of the model with  $(\gamma_X,\gamma_{\bar X})=(0,0)$ and to the $c=1$ state
with  $(\gamma_X,\gamma_{\bar X})=(\pi,0)$ or $(0,\pi)$.

Finally, we study the same model with impurities.
We can define  the Berry phase even in such a model by imposing the twisted boundary condition and 
by using the 
twist angle $(\phi_x,\phi_y)$ instead of the momentum $(k_x,k_y)$.
We then obtain $\gamma_x=0$ mod $2\pi$ for $\phi_y=0$. 
This is expected because the edge states that cross  
the zero energy at $k_y=0$ and $\pi$ in the pure model are no longer distinguished by $k_y$, and
hence $\gamma_x=0=\pi+\pi$ mod $2\pi$ is observed, even if the edge states remain.
Figure \ref{f:WTP}(g) shows the entanglement spectrum for $Y$ under the symmetric partition 
$n_Y=N/2$. 
At $\phi_y=0$ we see that
even with disorder, the four states seem  degenerate 
near the zero energy, which may originate from the zero-energy edge states of the pure model.
Here, the entanglement Berry phase for this state plays a crucial role 
when discussing the stability of the WT phase.
Let us calculate the entanglement Berry phase 
for a minimum subsystem $X$ with $n_X=1$ 
and its complement $\bar X$ with $n_{\bar X}=N-1$.
The spectrum of $X$ is displayed in Fig. \ref{f:WTP}(h). 
Note that it is indeed gapped, and the computed Berry phase is
$(\gamma_X,\gamma_{\bar X})=(\pi,\pi)$ even with disorder. 
These entanglement Berry phases imply that if one divides the system into two pieces $X$ and $\bar X$ 
and regard them as two one-dimensional chains, each subsystem has each edge states.
The natural sum rule $0=\gamma_x=\gamma_X+\gamma_{\bar X}=\pi+\pi$ mod $2\pi$ indeed holds.
In other words, a partitioning of the Berry phase enables us to observe the Berry phase $\pi$.

To summarize, we have argued that the entanglement topological numbers are invariant 
under the disentanglement of an entangled groundstate
and that they are topological numbers 
attached to disentangled  subsystems. 
In this sense, the entanglement topological numbers
serve as a partitioning of the topological numbers. 
We have introduced the entanglement Berry phase to show the stability of the edge states in the  WT phase.


We would like to thank K.-I. Imura and Y. Yoshimura for fruitful discussions on 
the Wilson-Dirac model.
This work was supported by Grants-in-Aid for Scientific Research (KAKENHI) Numbers 
25400388 (TF), 
25610101(YH), and 26247064
from Japan Society for the Promotion of Science (JSPS).


\begin{thebibliography}{31}
\expandafter\ifx\csname natexlab\endcsname\relax\def\natexlab#1{#1}\fi
\expandafter\ifx\csname bibnamefont\endcsname\relax
  \def\bibnamefont#1{#1}\fi
\expandafter\ifx\csname bibfnamefont\endcsname\relax
  \def\bibfnamefont#1{#1}\fi
\expandafter\ifx\csname citenamefont\endcsname\relax
  \def\citenamefont#1{#1}\fi
\expandafter\ifx\csname url\endcsname\relax
  \def\url#1{\texttt{#1}}\fi
\expandafter\ifx\csname urlprefix\endcsname\relax\def\urlprefix{URL }\fi
\providecommand{\bibinfo}[2]{#2}
\providecommand{\eprint}[2][]{\url{#2}}

\bibitem[{\citenamefont{Holzhey et~al.}(1994)\citenamefont{Holzhey, Larsen, and
  Wilczek}}]{Holzhey1994443}
\bibinfo{author}{\bibfnamefont{C.}~\bibnamefont{Holzhey}},
  \bibinfo{author}{\bibfnamefont{F.}~\bibnamefont{Larsen}}, \bibnamefont{and}
  \bibinfo{author}{\bibfnamefont{F.}~\bibnamefont{Wilczek}},
  \bibinfo{journal}{Nuclear Physics B} \textbf{\bibinfo{volume}{424}},
  \bibinfo{pages}{443 } (\bibinfo{year}{1994}), ISSN \bibinfo{issn}{0550-3213},
  \urlprefix\url{http://www.sciencedirect.com/science/article/pii/0550321394904022}.

\bibitem[{\citenamefont{Vidal et~al.}(2003)\citenamefont{Vidal, Latorre, Rico,
  and Kitaev}}]{PhysRevLett.90.227902}
\bibinfo{author}{\bibfnamefont{G.}~\bibnamefont{Vidal}},
  \bibinfo{author}{\bibfnamefont{J.~I.} \bibnamefont{Latorre}},
  \bibinfo{author}{\bibfnamefont{E.}~\bibnamefont{Rico}}, \bibnamefont{and}
  \bibinfo{author}{\bibfnamefont{A.}~\bibnamefont{Kitaev}},
  \bibinfo{journal}{Phys. Rev. Lett.} \textbf{\bibinfo{volume}{90}},
  \bibinfo{pages}{227902} (\bibinfo{year}{2003}),
  \urlprefix\url{http://link.aps.org/doi/10.1103/PhysRevLett.90.227902}.

\bibitem[{\citenamefont{Ryu and Hatsugai}(2006)}]{Ryu:2006fk}
\bibinfo{author}{\bibfnamefont{S.}~\bibnamefont{Ryu}} \bibnamefont{and}
  \bibinfo{author}{\bibfnamefont{Y.}~\bibnamefont{Hatsugai}},
  \bibinfo{journal}{Phys. Rev. B} \textbf{\bibinfo{volume}{73}},
  \bibinfo{pages}{245115} (\bibinfo{year}{2006}), \eprint{cond-mat/0601237},
  \urlprefix\url{http://arXiv.org/abs/cond-mat/0601237}.

\bibitem[{\citenamefont{Li and Haldane}(2008)}]{PhysRevLett.101.010504}
\bibinfo{author}{\bibfnamefont{H.}~\bibnamefont{Li}} \bibnamefont{and}
  \bibinfo{author}{\bibfnamefont{F.}~\bibnamefont{Haldane}},
  \bibinfo{journal}{Phys. Rev. Lett.} \textbf{\bibinfo{volume}{101}},
  \bibinfo{pages}{010504} (\bibinfo{year}{2008}),
  \urlprefix\url{http://link.aps.org/doi/10.1103/PhysRevLett.101.010504}.

\bibitem[{\citenamefont{Turner et~al.}(2010)\citenamefont{Turner, Zhang, and
  Vishwanath}}]{Turner:2010ys}
\bibinfo{author}{\bibfnamefont{A.~M.} \bibnamefont{Turner}},
  \bibinfo{author}{\bibfnamefont{Y.}~\bibnamefont{Zhang}}, \bibnamefont{and}
  \bibinfo{author}{\bibfnamefont{A.}~\bibnamefont{Vishwanath}},
  \bibinfo{journal}{Physical Review B} \textbf{\bibinfo{volume}{82}},
  \bibinfo{pages}{241102} (\bibinfo{year}{2010}).

\bibitem[{\citenamefont{Prodan et~al.}(2010)\citenamefont{Prodan, Hughes, and
  Bernevig}}]{Prodan:2010fj}
\bibinfo{author}{\bibfnamefont{E.}~\bibnamefont{Prodan}},
  \bibinfo{author}{\bibfnamefont{T.~L.} \bibnamefont{Hughes}},
  \bibnamefont{and} \bibinfo{author}{\bibfnamefont{B.~A.}
  \bibnamefont{Bernevig}}, \bibinfo{journal}{Physical Review Letters}
  \textbf{\bibinfo{volume}{105}}, \bibinfo{pages}{115501}
  (\bibinfo{year}{2010}),
  \urlprefix\url{http://link.aps.org/doi/10.1103/PhysRevLett.105.115501}.

\bibitem[{\citenamefont{Cirac et~al.}(2011)\citenamefont{Cirac, Poilblanc,
  Schuch, and Verstraete}}]{Cirac:2011ul}
\bibinfo{author}{\bibfnamefont{J.~I.} \bibnamefont{Cirac}},
  \bibinfo{author}{\bibfnamefont{D.}~\bibnamefont{Poilblanc}},
  \bibinfo{author}{\bibfnamefont{N.}~\bibnamefont{Schuch}}, \bibnamefont{and}
  \bibinfo{author}{\bibfnamefont{F.}~\bibnamefont{Verstraete}},
  \bibinfo{journal}{Physical Review B} \textbf{\bibinfo{volume}{83}},
  \bibinfo{pages}{245134} (\bibinfo{year}{2011}),
  \urlprefix\url{http://link.aps.org/doi/10.1103/PhysRevB.83.245134}.

\bibitem[{\citenamefont{Hughes et~al.}(2011)\citenamefont{Hughes, Prodan, and
  Bernevig}}]{Hughes:2011rm}
\bibinfo{author}{\bibfnamefont{T.~L.} \bibnamefont{Hughes}},
  \bibinfo{author}{\bibfnamefont{E.}~\bibnamefont{Prodan}}, \bibnamefont{and}
  \bibinfo{author}{\bibfnamefont{B.~A.} \bibnamefont{Bernevig}},
  \bibinfo{journal}{Physical Review B} \textbf{\bibinfo{volume}{83}},
  \bibinfo{pages}{245132} (\bibinfo{year}{2011}),
  \urlprefix\url{http://link.aps.org/doi/10.1103/PhysRevB.83.245132}.

\bibitem[{\citenamefont{Alexandradinata
  et~al.}(2011)\citenamefont{Alexandradinata, Hughes, and
  Bernevig}}]{Alexandradinata:2011gf}
\bibinfo{author}{\bibfnamefont{A.}~\bibnamefont{Alexandradinata}},
  \bibinfo{author}{\bibfnamefont{T.~L.} \bibnamefont{Hughes}},
  \bibnamefont{and} \bibinfo{author}{\bibfnamefont{B.~A.}
  \bibnamefont{Bernevig}}, \bibinfo{journal}{Physical Review B}
  \textbf{\bibinfo{volume}{84}}, \bibinfo{pages}{195103}
  (\bibinfo{year}{2011}),
  \urlprefix\url{http://link.aps.org/doi/10.1103/PhysRevB.84.195103}.

\bibitem[{\citenamefont{Huang and Arovas}(2012)}]{Huang:2012db}
\bibinfo{author}{\bibfnamefont{Z.}~\bibnamefont{Huang}} \bibnamefont{and}
  \bibinfo{author}{\bibfnamefont{D.~P.} \bibnamefont{Arovas}},
  \bibinfo{journal}{Physical Review B} \textbf{\bibinfo{volume}{86}},
  \bibinfo{pages}{245109} (\bibinfo{year}{2012}),
  \urlprefix\url{http://link.aps.org/doi/10.1103/PhysRevB.86.245109}.

\bibitem[{\citenamefont{Fang et~al.}(2013)\citenamefont{Fang, Gilbert, and
  Bernevig}}]{Fang:2013dq}
\bibinfo{author}{\bibfnamefont{C.}~\bibnamefont{Fang}},
  \bibinfo{author}{\bibfnamefont{M.~J.} \bibnamefont{Gilbert}},
  \bibnamefont{and} \bibinfo{author}{\bibfnamefont{B.~A.}
  \bibnamefont{Bernevig}}, \bibinfo{journal}{Physical Review B}
  \textbf{\bibinfo{volume}{87}}, \bibinfo{pages}{035119}
  (\bibinfo{year}{2013}),
  \urlprefix\url{http://link.aps.org/doi/10.1103/PhysRevB.87.035119}.

\bibitem[{\citenamefont{Lou et~al.}(2011)\citenamefont{Lou, Tanaka, Katsura,
  and Kawashima}}]{Lou:2011jk}
\bibinfo{author}{\bibfnamefont{J.}~\bibnamefont{Lou}},
  \bibinfo{author}{\bibfnamefont{S.}~\bibnamefont{Tanaka}},
  \bibinfo{author}{\bibfnamefont{H.}~\bibnamefont{Katsura}}, \bibnamefont{and}
  \bibinfo{author}{\bibfnamefont{N.}~\bibnamefont{Kawashima}},
  \bibinfo{journal}{Physical Review B} \textbf{\bibinfo{volume}{84}},
  \bibinfo{pages}{245128} (\bibinfo{year}{2011}),
  \urlprefix\url{http://link.aps.org/doi/10.1103/PhysRevB.84.245128}.

\bibitem[{\citenamefont{Tanaka et~al.}(2012)\citenamefont{Tanaka, Tamura, and
  Katsura}}]{Tanaka:2012qv}
\bibinfo{author}{\bibfnamefont{S.}~\bibnamefont{Tanaka}},
  \bibinfo{author}{\bibfnamefont{R.}~\bibnamefont{Tamura}}, \bibnamefont{and}
  \bibinfo{author}{\bibfnamefont{H.}~\bibnamefont{Katsura}},
  \bibinfo{journal}{Physical Review A} \textbf{\bibinfo{volume}{86}},
  \bibinfo{pages}{032326} (\bibinfo{year}{2012}),
  \urlprefix\url{http://link.aps.org/doi/10.1103/PhysRevA.86.032326}.

\bibitem[{\citenamefont{Hatsugai}(1993)}]{Hatsugai:1993fk}
\bibinfo{author}{\bibfnamefont{Y.}~\bibnamefont{Hatsugai}},
  \bibinfo{journal}{Physical Review Letters} \textbf{\bibinfo{volume}{71}},
  \bibinfo{pages}{3697} (\bibinfo{year}{1993}),
  \urlprefix\url{http://link.aps.org/doi/10.1103/PhysRevLett.71.3697}.

\bibitem[{\citenamefont{Hsieh and Fu}(2014)}]{Hsieh:2014qy}
\bibinfo{author}{\bibfnamefont{T.~H.} \bibnamefont{Hsieh}} \bibnamefont{and}
  \bibinfo{author}{\bibfnamefont{L.}~\bibnamefont{Fu}},
  \bibinfo{journal}{Physical Review Letters} \textbf{\bibinfo{volume}{113}},
  \bibinfo{pages}{106801} (\bibinfo{year}{2014}),
  \urlprefix\url{http://link.aps.org/doi/10.1103/PhysRevLett.113.106801}.

\bibitem[{\citenamefont{Hsieh et~al.}(2014)\citenamefont{Hsieh, Fu, and
  Qi}}]{Hsieh:2014jk}
\bibinfo{author}{\bibfnamefont{T.~H.} \bibnamefont{Hsieh}},
  \bibinfo{author}{\bibfnamefont{L.}~\bibnamefont{Fu}}, \bibnamefont{and}
  \bibinfo{author}{\bibfnamefont{X.-L.} \bibnamefont{Qi}},
  \bibinfo{journal}{Physical Review B} \textbf{\bibinfo{volume}{90}},
  \bibinfo{pages}{085137} (\bibinfo{year}{2014}),
  \urlprefix\url{http://link.aps.org/doi/10.1103/PhysRevB.90.085137}.

\bibitem[{\citenamefont{Fukui and Hatsugai}(2014)}]{Fukui:2014qv}
\bibinfo{author}{\bibfnamefont{T.}~\bibnamefont{Fukui}} \bibnamefont{and}
  \bibinfo{author}{\bibfnamefont{Y.}~\bibnamefont{Hatsugai}},
  \bibinfo{journal}{Journal of the Physical Society of Japan}
  \textbf{\bibinfo{volume}{83}}, \bibinfo{pages}{113705}
  (\bibinfo{year}{2014}),
  \urlprefix\url{http://dx.doi.org/10.7566/JPSJ.83.113705}.

\bibitem[{\citenamefont{Vijay and Fu}(2014)}]{vijay2014entanglement}
\bibinfo{author}{\bibfnamefont{S.}~\bibnamefont{Vijay}} \bibnamefont{and}
  \bibinfo{author}{\bibfnamefont{L.}~\bibnamefont{Fu}}, \bibinfo{journal}{arXiv
  preprint arXiv:1412.4733}  (\bibinfo{year}{2014}).

\bibitem[{\citenamefont{Fukui et~al.}(2013)\citenamefont{Fukui, Imura, and
  Hatsugai}}]{Fukui:2013mz}
\bibinfo{author}{\bibfnamefont{T.}~\bibnamefont{Fukui}},
  \bibinfo{author}{\bibfnamefont{K.-I.} \bibnamefont{Imura}}, \bibnamefont{and}
  \bibinfo{author}{\bibfnamefont{Y.}~\bibnamefont{Hatsugai}},
  \bibinfo{journal}{Journal of the Physical Society of Japan}
  \textbf{\bibinfo{volume}{82}}, \bibinfo{pages}{073708}
  (\bibinfo{year}{2013}),
  \urlprefix\url{http://dx.doi.org/10.7566/JPSJ.82.073708}.

\bibitem[{\citenamefont{Yoshimura et~al.}(2014)\citenamefont{Yoshimura, Imura,
  Fukui, and Hatsugai}}]{Yoshimura:2014qf}
\bibinfo{author}{\bibfnamefont{Y.}~\bibnamefont{Yoshimura}},
  \bibinfo{author}{\bibfnamefont{K.-I.} \bibnamefont{Imura}},
  \bibinfo{author}{\bibfnamefont{T.}~\bibnamefont{Fukui}}, \bibnamefont{and}
  \bibinfo{author}{\bibfnamefont{Y.}~\bibnamefont{Hatsugai}},
  \bibinfo{journal}{Physical Review B} \textbf{\bibinfo{volume}{90}},
  \bibinfo{pages}{155443} (\bibinfo{year}{2014}),
  \urlprefix\url{http://link.aps.org/doi/10.1103/PhysRevB.90.155443}.

\bibitem[{\citenamefont{Peschel}(2003)}]{Peschel:2003uq}
\bibinfo{author}{\bibfnamefont{I.}~\bibnamefont{Peschel}},
  \bibinfo{journal}{J.Phys.A: Math.Gen.} \textbf{\bibinfo{volume}{36}},
  \bibinfo{pages}{L205} (\bibinfo{year}{2003}), \eprint{cond-mat/0212631},
  \urlprefix\url{http://arXiv.org/abs/cond-mat/0212631}.

\bibitem[{\citenamefont{Kane and Mele}(2005)}]{Kane:2005fk}
\bibinfo{author}{\bibfnamefont{C.~L.} \bibnamefont{Kane}} \bibnamefont{and}
  \bibinfo{author}{\bibfnamefont{E.~J.} \bibnamefont{Mele}},
  \bibinfo{journal}{Physical Review Letters} \textbf{\bibinfo{volume}{95}}
  (\bibinfo{year}{2005}),
  \urlprefix\url{http://link.aps.org/doi/10.1103/PhysRevLett.95.146802}.

\bibitem[{\citenamefont{Sheng et~al.}(2006)\citenamefont{Sheng, Weng, Sheng,
  and Haldane}}]{PhysRevLett.97.036808}
\bibinfo{author}{\bibfnamefont{D.~N.} \bibnamefont{Sheng}},
  \bibinfo{author}{\bibfnamefont{Z.~Y.} \bibnamefont{Weng}},
  \bibinfo{author}{\bibfnamefont{L.}~\bibnamefont{Sheng}}, \bibnamefont{and}
  \bibinfo{author}{\bibfnamefont{F.~D.~M.} \bibnamefont{Haldane}},
  \bibinfo{journal}{Phys. Rev. Lett.} \textbf{\bibinfo{volume}{97}},
  \bibinfo{pages}{036808} (\bibinfo{year}{2006}),
  \urlprefix\url{http://link.aps.org/doi/10.1103/PhysRevLett.97.036808}.

\bibitem[{\citenamefont{Fukui and Hatsugai}(2007)}]{Fukui:2007sf}
\bibinfo{author}{\bibfnamefont{T.}~\bibnamefont{Fukui}} \bibnamefont{and}
  \bibinfo{author}{\bibfnamefont{Y.}~\bibnamefont{Hatsugai}},
  \bibinfo{journal}{Physical Review B} \textbf{\bibinfo{volume}{75}},
  \bibinfo{pages}{121403} (\bibinfo{year}{2007}),
  \urlprefix\url{http://link.aps.org/doi/10.1103/PhysRevB.75.121403}.

\bibitem[{\citenamefont{Ryu and Hatsugai}(2002)}]{Ryu:2002fk}
\bibinfo{author}{\bibfnamefont{S.}~\bibnamefont{Ryu}} \bibnamefont{and}
  \bibinfo{author}{\bibfnamefont{Y.}~\bibnamefont{Hatsugai}},
  \bibinfo{journal}{Physical Review Letters} \textbf{\bibinfo{volume}{89}},
  \bibinfo{pages}{077002} (\bibinfo{year}{2002}).

\bibitem[{\citenamefont{Hatsugai}(2006)}]{doi:10.1143/JPSJ.75.123601}
\bibinfo{author}{\bibfnamefont{Y.}~\bibnamefont{Hatsugai}},
  \bibinfo{journal}{Journal of the Physical Society of Japan}
  \textbf{\bibinfo{volume}{75}}, \bibinfo{pages}{123601}
  (\bibinfo{year}{2006}), \eprint{http://dx.doi.org/10.1143/JPSJ.75.123601},
  \urlprefix\url{http://dx.doi.org/10.1143/JPSJ.75.123601}.

\bibitem[{\citenamefont{Bernevig et~al.}(2006)\citenamefont{Bernevig, Hughes,
  and Zhang}}]{Bernevig:2006fk}
\bibinfo{author}{\bibfnamefont{B.~A.} \bibnamefont{Bernevig}},
  \bibinfo{author}{\bibfnamefont{T.~L.} \bibnamefont{Hughes}},
  \bibnamefont{and} \bibinfo{author}{\bibfnamefont{S.-C.} \bibnamefont{Zhang}},
  \bibinfo{journal}{Science,} \textbf{\bibinfo{volume}{314}},
  \bibinfo{pages}{1757} (\bibinfo{year}{2006}), \eprint{cond-mat/0611399},
  \urlprefix\url{http://arXiv.org/abs/cond-mat/0611399}.

\bibitem[{\citenamefont{Qi et~al.}(2008)\citenamefont{Qi, Hughes, and
  Zhang}}]{Qi:2008fk}
\bibinfo{author}{\bibfnamefont{X.-L.} \bibnamefont{Qi}},
  \bibinfo{author}{\bibfnamefont{T.~L.} \bibnamefont{Hughes}},
  \bibnamefont{and} \bibinfo{author}{\bibfnamefont{S.-C.} \bibnamefont{Zhang}},
  \bibinfo{journal}{Physical Review B} \textbf{\bibinfo{volume}{78}}
  (\bibinfo{year}{2008}),
  \urlprefix\url{http://link.aps.org/doi/10.1103/PhysRevB.78.195424}.

\bibitem[{\citenamefont{Jiang et~al.}(2014)\citenamefont{Jiang, Liu, Feng, Sun,
  and Xie}}]{Jiang:2014fk}
\bibinfo{author}{\bibfnamefont{H.}~\bibnamefont{Jiang}},
  \bibinfo{author}{\bibfnamefont{H.}~\bibnamefont{Liu}},
  \bibinfo{author}{\bibfnamefont{J.}~\bibnamefont{Feng}},
  \bibinfo{author}{\bibfnamefont{Q.}~\bibnamefont{Sun}}, \bibnamefont{and}
  \bibinfo{author}{\bibfnamefont{X.~C.} \bibnamefont{Xie}},
  \bibinfo{journal}{Physical Review Letters} \textbf{\bibinfo{volume}{112}},
  \bibinfo{pages}{176601} (\bibinfo{year}{2014}),
  \urlprefix\url{http://link.aps.org/doi/10.1103/PhysRevLett.112.176601}.

\bibitem[{\citenamefont{Diez et~al.}(2014{\natexlab{a}})\citenamefont{Diez,
  Fulga, Pikulin, Tworzyd{\l}o, and Beenakker}}]{Diez:2014xy}
\bibinfo{author}{\bibfnamefont{M.}~\bibnamefont{Diez}},
  \bibinfo{author}{\bibfnamefont{I.~C.} \bibnamefont{Fulga}},
  \bibinfo{author}{\bibfnamefont{D.~I.} \bibnamefont{Pikulin}},
  \bibinfo{author}{\bibfnamefont{J.}~\bibnamefont{Tworzyd{\l}o}},
  \bibnamefont{and} \bibinfo{author}{\bibfnamefont{C.~W.~J.}
  \bibnamefont{Beenakker}}, \bibinfo{journal}{New Journal of Physics}
  \textbf{\bibinfo{volume}{16}}, \bibinfo{pages}{063049}
  (\bibinfo{year}{2014}{\natexlab{a}}),
  \urlprefix\url{http://stacks.iop.org/1367-2630/16/i=6/a=063049}.

\bibitem[{\citenamefont{Diez et~al.}(2014{\natexlab{b}})\citenamefont{Diez,
  Pikulin, Fulga, and Tworzydlo}}]{Diez:2014nr}
\bibinfo{author}{\bibfnamefont{M.}~\bibnamefont{Diez}},
  \bibinfo{author}{\bibfnamefont{D.}~\bibnamefont{Pikulin}},
  \bibinfo{author}{\bibfnamefont{I.}~\bibnamefont{Fulga}}, \bibnamefont{and}
  \bibinfo{author}{\bibfnamefont{J.}~\bibnamefont{Tworzydlo}},
  \bibinfo{journal}{arXiv preprint arXiv:1412.3014}
  (\bibinfo{year}{2014}{\natexlab{b}}).

\end{thebibliography}

\end{document}